\documentclass[conference]{IEEEtran}

\usepackage{amsmath,dsfont,bbm,epsfig,amssymb,amsfonts,amstext,verbatim,amsopn,cite}
\usepackage{subfigure,multirow,multicol,lipsum,xfrac}
\usepackage{amsthm,ulem}
\usepackage{mathtools,amsthm}
\usepackage{perpage}
\usepackage{balance}
\usepackage{url}
\usepackage{amsfonts}
\usepackage{epsfig}
\usepackage[font={small}]{caption}
\usepackage{mleftright}
\mleftright
\usepackage{amsfonts,amsthm,xcolor,bbm}
\usepackage{bookmark}
\usepackage{tikz}
\usetikzlibrary{angles,quotes,arrows,decorations.pathmorphing}

\newcommand{\gruen}[1]{{\color{green!70!black}#1}}
\newcommand{\rot}[1]{{\color{red!85!black}#1}}
\def\expect{\mathop{\mbox{$\mathsf{E}$}}}

\usepackage{etoolbox}
\usepackage{algorithmicx}
\usepackage[Algorithm,ruled]{algorithm}
\usepackage{algpseudocode}
\usepackage{pifont}
\usepackage[utf8]{inputenc}
\usepackage[T1]{fontenc}  
\usepackage[nolist]{acronym}
\MakePerPage{footnote}
\usepackage{textcomp}
\usepackage{paralist}
\usepackage{enumitem}
\usepackage{bbm}
\usepackage[process=auto]{pstool}
\usepackage{tikz,pgfplots}
\usetikzlibrary{shapes,arrows}

\newcommand{\setR}{\mathbb{R}}

\newcommand{\setZ}{\mathbb{Z}}

\newcommand{\bz}{{\mathbf{z}}}

\newcommand{\br}{{\mathbf{r}}}
\newcommand{\bt}{{\mathbf{t}}}
\newcommand{\btan}{{\mathbf{z}_{\text t}}}
\newcommand{\brad}{{\mathbf{z}_{\text r}}}

\newcommand{\bm}{{\boldsymbol{m}}}

\newcommand{\bc}{{\boldsymbol{c}}}
\newcommand{\bw}{{\boldsymbol{w}}}

\renewcommand{\emph}[1]{\textit{#1}}

\begin{document}
\title{On Approximation, Bounding \& Exact Calculation of Average Block Error Probability for Random Code Ensembles}
%
\author{
\IEEEauthorblockN{
Ralf R. M\"uller\IEEEauthorrefmark{1}
}
\IEEEauthorblockA{
\IEEEauthorrefmark{1}Institute for Digital Communications, Friedrich-Alexander Universit\"at Erlangen-N\"urnberg, Germany\\
ralf.r.mueller@fau.de
}
}
%
%
\IEEEoverridecommandlockouts

\maketitle


\begin{abstract}
This paper presents a method to calculate the exact average block error probability of some random code ensembles under maximum-likelihood decoding. The proposed method is applicable to various channels and ensembles. The focus is on both spherical and Gaussian random codes on the additive white Gaussian noise channel as well as binary random codes on both the binary symmetric channel and the binary erasure channel. 

While for the uniform spherical ensemble Shannon, in 1959, argued with solid angles in $N$-dimensional space, the presented approach projects the problem into two dimensions and applies standard trigonometry. This simplifies the derivation and also allows for the analysis of the independent identically distributed (i.i.d.) Gaussian ensemble which turns out to perform better for short blocklengths and high rates. Moreover, a new lower bound on the average block error probability of the uniform spherical ensemble is found. For codes with more than three codewords, it is tighter than the sphere packing bound, but requires exactly the same computing effort. Furthermore, tight approximations are proposed to simplify the computation of both the exact average error probability and the two bounds. 

For the binary symmetric channel and the binary erasure channel, bounds on the average block error probability for i.i.d.\ random coding are derived and compared to the exact calculations.
\end{abstract}

\IEEEpeerreviewmaketitle

\section{Introduction}
\label{sec:intro}

Random codes were originally introduced as a tool to prove upper bounds on the block error probability of optimal codes \cite{shannon:48ab,shannon:59,gallager:65}.
With the advent of modern coding theory \cite{richardson:08}, random code constructions became the method of choice in many commercial applications. As a result, random codes emerged from an auxiliary means to study optimal codes to an object of primary interest.
Following this development, the focus of this paper is to better understand random code ensembles and their (ensemble-averaged) block error probability. 

In his 1959 seminal paper, Shannon \cite{shannon:59} calculates the exact block error probability of a spherical random code ensemble on the additive white Gaussian noise (AWGN) channel. However, the exact formula turned out to be hopeless to evaluate numerically given the computing tools of that time. Still he could find accurate bounds on the exact error probability even for optimal, not necessarily random codes. 
His lower bound is based on an argument resorting to fill the entire $N$-dimensional hypersphere with hyperspheres of smaller radii and commonly referred to as the {\it 1959 sphere packing bound}.
This bound and related ones were not easy to evaluate back in the 1960s \cite{slepian:63}. Subsequent works focussed on applying the sphere packing idea to a broader range of codes and channels \cite{shannon:67,dolinar:98a,
salema:02,valembois:04, wiechman:08}; see \cite{sason:06} for a tutorial. Improved computing power has made the 1959 sphere packing bound a very fast and useful tool for checking code performance \cite{vialle:99,pretty_good_codes}.

More recently, Polyanski et al.\ \cite{polyanskiy:10} derive bounds on the block error probability of random code ensembles which are reasonably tight even for blocklengths in the range of hundreds. Reference \cite{erseghe:16} shows how to compute these bounds more efficiently by means of Laplace transforms.
Other recent work targets mismatched decoding \cite{scarlett:14}, exact asymptotics \cite{honda:15,honda:17}, the log-volume of optimal codes \cite{moulin:17}, and further refinements of existing bounds \cite{altug:14a,altug:14b}.

The main contributions of this paper are as follows: 
\begin{itemize}
\item
It solves the problem with numerical evaluation of Shannon's 1959 formula for the average block error probability of the uniform spherical ensemble over a wide range of blocklengths. 
\item
It gives an alternative derivation of this block error probability without resorting to solid angles that is not limited to spherical codes, but more general. 
\item
It derives the exact average error probability for the independent identically distributed (i.i.d.) Gaussian ensemble on the AWGN channel.
\item 
It shows how to numerically evaluate known formulas for exact average block error probability of binary random coding on both the binary symmetric channel (BSC) and the binary erasure channel (BEC). 
\item
It finds a new lower bound on the average block error probability of the uniform spherical ensemble that is tighter than the 1959 sphere packing bound whenever the code contains at least four codewords.
\end{itemize}

The paper is organized as follows:
In Section~\ref{theory}, the extreme value methods is discussed as a means to calculate the average block error probability of random code ensembles. 
This method is applied to the uniform spherical ensemble and the i.i.d.\ Gaussian ensemble in Sections~\ref{sphrancod} and \ref{grc}, respectively. BSC and BEC are treated in Sections~\ref{bsc} and \ref{bec}, respectively.
Section~\ref{NumRes} discusses issues that arise from the numerical evaluation of the formulas in the previous sections. 
Section~\ref{conc} summarizes the conclusions.
\section{The Extreme Value Method}
\label{extvalmet}
\label{theory}
The general idea of this paper is to calculate the ensemble averaged block error probability of a random code by means of order statistics.

Consider a communication channel with transmitted codeword $\bt$ and received word $\br$ canonically described by $\Pr(\br|\bt)$.
Any decoder, even a mismatched and/or suboptimal one, uses a metric $m(\bw,\br)$ which scores any potential codeword $\bw$ given the received word $\br$. 
The codeword with the best score, i.e., the minimal metric, succeeds. 

For some channels and metrics, the best score is unique with probability 1 (almost surely). In such cases, the extreme value method may directly find exact results on the ensemble averaged block error probability.  If the best score is not unique, further refinements are needed to evaluate the probability of guessing correctly among several codewords with equal metrics.

Given that $\bc=\bw_{\tilde \imath}$ is the correct codeword out of a total of $M$ codewords and the best score is almost surely unique, the ensemble averaged block error probability is given by
\begin{align}
P_{\text e} &= \expect\limits_{\br} \expect\limits_{\bw_1,\dots,\bw_M|\br}
\Pr \left( \bigcup \limits_{i\ne \tilde\imath } \{ m(\bw_i, \br) < m(\bc,\br) \} \right) \\
&= \expect\limits_{\br} \expect\limits_{\bw_1,\dots,\bw_M|\br}\Pr\left(\min_{i\ne\tilde\imath} m(\bw_i,\br) < m(\bc,\br) \right).
\label{eq1}
\end{align}
Assuming that all codewords are generated statistically independent at random, the inner (conditional) expectation can be written as
\begin{equation}
\label{eq2}
1-\expect\limits_{\bc|\br} \left[1-{\text P}_{m|\br}\left(m(\bc,\br) \right)\right]^{M-1}
\end{equation} 
with ${\text P}_{\bm|\br}(\cdot)$ denoting the cumulative distribution function (CDF) of the metric of a {\it wrong} codeword conditioned on the received word $\br$.


The challenges in the extreme value method are as follows:
\begin{enumerate}
\item
Find an expression for the conditional CDF of the metric of a wrong codeword. This is not trivial for many elementary channels and decoders.
\item
Find a way to numerically calculate the right tail of the conditional CDF, which is very close to unity, raised to an exponentially large power (the number of codewords) with sufficient accuracy. 
\item
Calculate the metric of the correct codeword $m(\bc,\br)$.
\item
Evaluate the joint expectation over the received word and the correct codeword $\expect_{\br,\bc} = \expect_{\br} \expect_{\bc|\br}$.
\end{enumerate}
Some of these challenges are prohibitive, unless the channel, the random code ensemble, and the decoder metric have suitable properties. In subsequent sections, we give four examples of cases for which the extreme value method happens to work.

In all these examples, we utilize some invariance of the code ensemble and the channel. This is the rotational invariance of the uniform spherical and the i.i.d.\ Gaussian ensemble on the AWGN channel and the permutational invariance of i.i.d.\ random codes on the BSC and the BEC.
These invariances imply that the conditional CDF does not depend on all $N$ components of the received word $\br$, but only on a small subset of them.
For the uniform spherical ensemble on the AWGN channel and the i.i.d.\ ensemble on both the BSC and the BEC, this subset even is the empty set.
For the i.i.d.\ Gaussian ensemble, it is the radial component of the received word.
Without such helpful invariances, the evaluation of the joint expectation becomes a multi-dimensional integral whose dimensionality scales with the blocklength.
\section{Uniform Spherical Ensemble}
\label{sphrancod}
\begin{figure}[t]
\centering
\begin{tikzpicture}[
  >=stealth',
  pos=0.8,
  photon/.style={decorate,decoration={snake,post length=4mm}}
]
\coordinate (C) at (3.1556,-0.5315);
\coordinate (Z) at (3.9444,-0.6644);
\coordinate (O) at (0,0);
\coordinate (R) at (4.4427,2.294);
\coordinate (W) at (-2.7426,2.0749);
\coordinate (P) at (-1.3194,-0.6813);
\gruen{\draw[dashed,thick,->] (O) -- (P) node[pos=.5, sloped, above] {$\boldsymbol{w}_{{\text r},i}$};}
\gruen{\draw[dashed,thick,->] (P) -- (W) node[pos=.4, sloped, above] {$\boldsymbol{w}_{{\text t},i}$};}
\draw[thick,blue,->] (O) -- (C) node[pos=.65, sloped, above] {$\boldsymbol{c}$};
\rot{\draw[thick,->] (C)-- (Z) node[midway,sloped,above] {$\mathbf{z}_{\rm r}$};
\draw[thick,->] (Z) -- (R) node[midway,sloped,above] {$\mathbf{z}_{\rm t}$};}
{\draw[thick,->] (O) -- (R) node[midway,sloped,above] {$\mathbf{r}$};}
\gruen{\draw[thick,->] (O) -- (W) node[pos=.65, sloped, above] {$\boldsymbol{w}_i$};}
\draw[thick,blue,<-] pic ["$\beta$"scale=1, draw, thick, angle eccentricity=1.8] {angle = C--O--R};
\gruen{\draw[thick,->] pic ["$\alpha_i$"scale=1, draw, thick, angle eccentricity=1.8] {angle = R--O--W};}
\end{tikzpicture}
\caption{Definition of angles and decomposition into radial and tangential components.}
\label{triangle}
\end{figure}
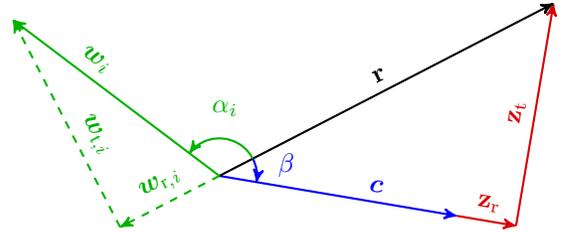

Consider a real-valued AWGN channel. The random codebook of size $2^{NR}$ is chosen from statistically independent points uniformly distributed on an $N$-dimensional hypersphere of radius $\sqrt{NP}$.
The transmitted codeword is denoted by $\bc$. 
It is distorted by independent AWGN of zero mean and variance $1$ denoted by $\bz$.
We decompose the noise vector into a radial component $\brad$ which is collinear to the code word $\bc$ and a tangential component $\btan$ which is orthogonal to the code word $\bc$, cf.\ Figure \ref{triangle}.
Note that such a decomposition of an i.i.d.\ Gaussian vector leads to two new Gaussian vectors which have independent components in appropriately chosen Cartesian coordinates.
We denote the angle between the codeword $\bc$ and the received word $\br$ as $\beta$ and the angle between an alternative codeword $\bw_i$ and the received word as $\alpha_i$.  

\subsection{Conditional CDF}
All codewords are uniformly distributed on the hypersphere. Their joint distribution is thus invariant to any rotation around the origin. We utilize this property and rotate the hypersphere in such a way that the received word becomes collinear to the first unit vector of a Cartesian coordinate system. 

The correlation coefficient between a codeword and the received word scores the likelihood of that codeword. Unlike the metric defined in Section~\ref{extvalmet}, it is the larger, the more likely a codeword is.

A uniform distribution on the unit hypersphere can be generated by normalizing an i.i.d.\ Gaussian random vector to unit length.
Denoting 
\begin{equation}
\label{defrho}
\rho_i = \cos\alpha_i,
\end{equation}
we can construct the squared correlation coefficient between the received word and the $i^{\text{th}}$ codeword out of $N$ i.i.d.\ Gaussian random variables $g_{i,1},\dots,g_{i,N}$ with zero mean and unit variance as
\begin{equation}
\label{rayleigh}
{\rho}_i^2 = {{g_{i,1}^2} }{\Big/ {\sum\limits_{n=1}^N g_{i,n}^2}}.
\end{equation}
In the numerator, only the first Gaussian random variable shows up due to the inner product with the first unit vector of the coordinate system.
The ratio in \eqref{rayleigh} is known to be distributed according to the beta distribution with shape parameters $\frac12$ and $\frac {N-1}2$. The corresponding density is given by \cite{casella:02}
\begin{equation}
{\text p}_{\rho^2}(r) = \frac1{{\rm B}(\frac12,\frac {N-1}2)}\,  r^{-\frac12} (1-r)^{\frac {N-3}2}.
\end{equation}
with $B(\cdot,\cdot)$ denoting the beta function.
Substituting ${\rho}=\sqrt r$ leads to the density
\begin{equation}
{\text p}_{|\rho|}(\rho) = \frac 2{{\rm B}(\frac12,\frac {N-1}2)} \, (1-\rho^2)^{\frac {N-3}2}, \quad \rho\in [0,1].
\end{equation}
By symmetry around $\rho=0$, it is straightforward to show that
\begin{equation}
\label{pdfrho}
{\text p}_{\rho}(\rho) = \frac 1{{\rm B}(\frac12,\frac {N-1}2)} \, (1-\rho^2)^{\frac {N-3}2},\quad \rho \in [-1,+1].
\end{equation}
This implies that the conditional CDF of 
\begin{equation}
\label{rhomax}
\varrho = \max\limits_i \rho_i
\end{equation}
is given by
\begin{align}
{\text P}_{\varrho|\br}(\varrho) &= \left[
 \int\limits_{-1}^{\varrho} {\text p}_{\rho}(\rho) {\text d}\rho\right]^{2^{NR}-1} 
\end{align}
since there are $2^{NR}-1$ alternative codewords and ${\text p}_\rho(\cdot)={\text p}_{\rho|\br}(\cdot)$ due to the rotational invariance of the code construction.

Defining the lower regularized incomplete beta function
\begin{equation}
\label{defbetainc}
{\rm B}(a,b,x) = \frac1{{\rm B}(a,b)} \int\limits^x_0 \xi^{a-1} (1-\xi)^{b-1} {\rm d}\xi,
\end{equation}
and substituting $x=\varrho^2$,
we get
\begin{align}
{\text P}_{\varrho|\br}(\varrho) 
&=\left[\frac12 + \frac{{\rm sign} \varrho}2\,{\text B}\left(\frac12,\frac {N-1}2,\varrho^2\right)  \right]^{2^{NR}-1}.
\label{rhodis}
\end{align}

\subsection{Metric of the Correct Codeword}
Consider the right triangle formed by the received word $\br$, the tangential noise component $\btan$, and the sum of codeword and radial noise component $\bc+\brad$ in Figure \ref{triangle}.
Denoting
\begin{equation}
\label{defchi}
\chi = \|\btan\|, 
\end{equation}
we recognize that $\chi^2$ follows a chi-square distribution with $N-1$ degrees of freedom and probability density
\begin{equation}
{\text p}_{\chi^2}(x) = \frac{x^{\frac{N-3}2} {\rm e}^{-\frac x2}}
{2^{\frac{N-1}2} \Gamma\left(\frac{N-1}2\right)}.
\label{chisden}
\end{equation} 
Mean and variance of the tangential noise $\chi$ are
\begin{align}
\label{meanchi}
\mu_{\chi} &= \frac{\sqrt 2\Gamma(\frac N2)}{\Gamma(\frac{N-1}2)} <\sqrt{N-1} \\
\label{varchi}
\sigma_{\chi}^2 &= N-1- \frac{2\Gamma^2(\frac N2)}{\Gamma^2(\frac{N-1}2)} < \frac 12
\end{align}
as can be readily shown by integration over \eqref{chisden}.

The radial component of the noise is Gaussian distributed with zero mean and unit variance and will be denoted by $z_{\text r}$.
It is collinear to the codeword. The sum of the two is denoted as
\begin{equation}
s=\sqrt {NP} + z_{\text r}.
\end{equation}
and Gaussian distributed with mean $\sqrt{NP}$ and unit variance.
Comparing to \eqref{varchi}, it is remarkable that the radial noise has more than twice the variance of the tangential noise.

The angle to the true codeword can be expressed as 
\begin{equation}
\label{deftan}
\cos\beta = \frac{s}{\sqrt{ s^2+ \chi^2}}
\end{equation}
by standard trigonometric considerations, cf.\ Figure \ref{triangle}.
Note that due to the joint independence of all codewords and noise, the angles $\alpha_i$ and $\beta$ are statistically independent.

\subsection{Joint Expectation}
The exact block error probability is now found collecting previous results. In particular, we use \eqref{eq1}, \eqref{eq2}, and \eqref{deftan}  to get
\begin{align}
P_{\text e} 
=1-\int\limits_0^\infty\!\! \int\limits_{\setR} {\text P}_{\varrho|\br}\left( {\frac{s}{\sqrt{s^2+x}}}\right) \frac{ {\text e}^{-\frac{\left(s-\sqrt{ NP}\right)^2}2}}{\sqrt{2\pi}} \,{\text d}s\, {\text p}_{\chi^2}(x) \,{\text d}x 
\label{eq25n}
\end{align}
with ${\text P}_{\varrho|\br}(\cdot)$ and ${\text p}_{\chi^2}(\cdot)$ specified in \eqref{rhodis} and \eqref{chisden}, respectively.

Following \cite{shannon:59}, the distribution of the ratio 
\begin{equation}
\label{deft}
t= \frac s\chi \sqrt{N-1}
\end{equation}
is identified as a noncentral $t$-distribution with $N-1$ degrees of freedom and noncentrality parameter $\sqrt{NP}$. 
Its density will be denoted by ${\text p}_t(t,N-1,\sqrt{NP})$, in the sequel.
This way, the double integral in \eqref{eq25n} can be simplified to a single one
\begin{align}
\label{jumpdef}
P_{\text e}=1-\int\limits_{\setR} {\text P}_{\varrho|\br}\left( {\frac{t}{\sqrt{t^2+N-1}}}\right) {\text p}_{t}(t) {\rm d}t. 
\end{align}
This expression is known from \cite{shannon:59}. Given the numerical computing standards of that time, it was impossible to evaluate it for relevant blocklengths. Even nowadays, the numerical evaluation is considered very challenging. Section~\ref{numres}, shows how to evaluate it for almost any blocklength within milliseconds on a desktop computer.

\subsection{The Median Lower Bound}
\label{lowerbounds}

While the average block error probability of the uniform spherical ensemble can be calculated exactly, some applications prefer calculation speed over accuracy. In the sequel, we present a novel tight lower bound on the average block error probability of this ensemble and discuss its connection to the well-known sphere packing bound.

Let us replace the exact shape of ${\text P}_{\varrho|\br}(\cdot)$ by a unit step function at its median value $m_\varrho$. This reduces the error probability, as we only shift probability for close wrong codewords to more distant wrong codewords. Clearly, the close wrong codewords can be reached more easily by the noise than the distant ones.

Due to \eqref{jumpdef}, the jump occurs at
\begin{equation}
m_\varrho = \frac{t}{\sqrt{t^2+N-1}}.
\end{equation}
Solving for $t$, the error probability is lower bounded by
\begin{equation}
\label{spb}
P_{\text e} > 
{\text P}_t\left(m_\varrho \sqrt{\frac{N-1}{1-m_\varrho^2}},N-1,\sqrt{NP}\right)
\end{equation}
with ${\text P}_t(t,N-1,\sqrt{NP})$ denoting the cumulative distribution function of the noncentral $t$-distribution. 

The median value $m_\varrho$ is easily found by setting the cumulative distribution function \eqref{rhodis} to $\frac12$.
Since the median $m_\varrho$ is clearly positive, we obtain
\begin{align}
\label{mrho}
m_\varrho^2 &= {\text B}^{-1} \left(\frac 12,\frac {N-1}2, 2^{1-({2^{NR} -1})^{-1}} - 1\right) \\
& =1- {\text B}^{-1} \left(\frac {N-1}2, \frac12, 2-2^{1-({2^{NR} -1})^{-1}} \right)\\
&> 1-{\text B}^{-1} \left(\frac {N-1}2, \frac12,  \frac {2\ln 2}{2^{NR} -1}\right)
\label{ewb}
\end{align}
where ${\text B}^{-1}(a,b,x)$ denotes the inverse with respect to composition of ${\text B}(a,b,x)$. 
The inequality follows from the first order Taylor series of the exponential function
\begin{equation}
2^{-x} > 1-x \ln 2 \qquad \mbox{for} \qquad x = ({2^{NR} -1})^{-1}.
\end{equation}
Inequality \eqref{ewb} is useful as \eqref{mrho} can numerically only be evaluated for moderate values of $NR$.

\newcommand{\rsp}{\rho_{\text{SP}}}
Another tight lower bound utilizing the fast hardening of the conditional CDF is Shannon's 1959 sphere packing bound.
In \cite{shannon:59}, he derives it by arguing that if all Voronoi regions would be equal in size and circular, error probability would be improved. 
Thus, the total solid angle should be $2^{NR}$ times the solid angle of one circular Voronoi region.
In the view of this article, his argument translates into the following line of thought: 
The edge of the circular Voronoi region is such that the probability of being within it is $2^{-NR}$.
Specifying the edge of the circular Voronoi region by the cosine of the angle between edge and center,
we get with \eqref{defrho} and \eqref{pdfrho}
\begin{align}
{2^{-NR}} = \int\limits^{1}_{\rsp} p_\rho(\rho){\text d}\rho
 & = \frac12 - \frac12 \,{\text B}\left(\frac12,\frac{N-1}2, \rsp^2 \right)\\
&=  \frac12 \,{\text B}\left(\frac{N-1}2, \frac12, 1-\rsp^2 \right)
\end{align}
and
\begin{align}
\label{spb36}
\rsp^2  = 1- {\text B}^{-1} \left(\frac {N-1}2, \frac12,  \frac {2}{2^{NR} }\right).
\end{align}
Interestingly, substituting $\rsp$ for $m_\varrho$ in \eqref{spb} gives exactly the sphere packing bound found in \cite{shannon:59}.

Note that for any $x\ge 2$,
$
2^x -1 > 2^x \ln 2.
$
Thus, whenever the code contains at least 4 codewords, the sphere packing bound is less tight than the median bound \eqref{ewb}. Unless $NR$ is small, the number of codewords is much larger than 1.
In that regime, the sphere packing assumption, i.e., the assumption that the Voronoi regions are hyperspheres, is equivalent to having $1/\ln 2\approx 1.443$ times more codewords available at the same given minimum distance.

Unlike the 1959 sphere packing bound, the median bound only applies to the {\it average} block error probability of the uniform spherical ensemble. The former one even bounds the minimum block error probability of spherical codes, regardless of whether they are random or deterministic.

\section{I.i.d.\ Gaussian Ensemble}
\label{grc}
Consider now the case that the codewords are not on the hypersphere, but their components are i.i.d.\ Gaussian with zero mean and variance $P$.
If one normalized these codewords to Euclidean norm $\sqrt{NP}$, we would be back at the uniform spherical ensemble. 
The additional radial component of the codewords provides an additional dimension in signal space.
As shown in Section~\ref{numresIID}, the IID\ Gaussian ensemble is superior to the uniform spherical ensemble for very short blocklengths. It is also useful to analyze the Gaussian multiple-access channel at finite blocklength \cite{mueller:20b}.

We decompose the alternative codeword $\bw_i\ne \bc$ into a radial and a tangential component relative to the received word $\br$ as in Figure \ref{triangle}
\begin{equation}
\bw_i = \bw_{\text t,i} + \bw_{\text r,i}.
\end{equation}
Note that $\|\bw_{\text t,i}\|^2/P$ is chi-squared distributed with $N-1$ degrees of freedom. Its probability density is given by \eqref{chisden}.
The radial component $w_{\text r,i}$ is zero mean Gaussian with variance $P$.
The two components are statistically independent.
The normalized squared distance to the received word \begin{equation}
d_i^2 = \frac{\|\bw_{\text t,i}\|^2 + \left( \|\br\|-w_{\text r,i}\right)^2}P
\end{equation}
serves as metric for maximum-likelihood decoding.
Conditioned on the received word $\bf r$, it follows a noncentral chi-square distribution with $N$ degrees of freedom, noncentrality parameter $\|\br\|^2/P$, and cumulative distribution function  \cite{casella:02}
\begin{equation}
\label{ncx2cdf}
{\text P}_{d^2|\br}(d) = 1 -{\text Q}_{\frac N2}\left(\|\br\|/\sqrt P,\sqrt d\right)
\end{equation}
with ${\text Q}_M(a,b)$ denoting the generalized Marcum Q-function.

The squared distance of the received word $\br$ to the correct codeword $\bc$ is 
\begin{equation}
\|\bz\|^2 = \chi^2 + z_{\text r}^2
\end{equation}
with Figure~\ref{triangle} and \eqref{defchi}.
Furthermore,
\begin{equation}
\|\br\|^2 =  \chi^2 + (z_{\text r} + \|\bc\|)^2.
\end{equation}
Collecting these results, we get the error probability with \eqref{eq1} and \eqref{eq2} as
\begin{align}
P_{\text e} 
 & = 1- \expect\limits_{\br,\bz} \left[{\text Q}_{\frac N2}\left(\frac{\|\br\|}{\sqrt P},\frac{\|\bz\|}{\sqrt P} \right) \right]^{2^{NR}-1}\\
 & = 1- \expect\limits_{\|\bc\|,z_{\text r}, \chi} \left[{\text Q}_{\frac N2}\left( \scriptstyle \sqrt{\frac{\chi^2 + (\|\bc\| +z_{\text r})^2}{P}},\sqrt{\frac{\chi^2 +z_{\text r}^2}P}  \right) \right]^{2^{NR}-1}.
 \label{gaussfinal}
\end{align}
The random variables $\|\bc\|$, ${z_{\text r}}$, and $\chi$ are statistically independent and their distributions are known. The expectation in \eqref{gaussfinal}, however, can only be evaluated numerically. 

Technically, the median bound is not restricted to the uniform spherical ensemble, but can be calculated for any ensemble where the functional inverse of the conditional CDF can be found.
For the i.i.d.\ Gaussian ensemble, this requires the inverse of the generalized Marcum Q-function. Its calculation is not straightforward \cite{helstrom:98,gil:14}. The author could not obtain significant savings of computation time over the exact calculation via \eqref{gaussfinal}.

\section{Binary Symmetric Channel}
\label{bsc}

Consider now a BSC with i.i.d.\ binary codewords having equal probability.
In order to streamline the presentation with respect to the previous sections, let the alphabet of the code be $\{+1,-1\}$. 
Since all codewords and the received word lie on a hypersphere in $N$ dimensions, we can argue in the same way as for spherical random coding on the AWGN channel: No error occurs, as long as the smallest angle between the received word and a wrong codeword is larger than the angle between the received word and the true codeword.

The correlation coefficient between a random wrong codeword and the received word scaled by the blocklength 
\begin{equation}
\label{antio}
o_i = N \cos\alpha_ i
\end{equation} 
is called {\it antipodal overlap} in the sequel and defined in analogy to \eqref{defrho}. 
In order to simplify subsequent notation, we also introduce the {\it Boolean overlap}
\begin{equation}
l_i = \frac{N+o_i}2 
\end{equation}
which corresponds to codewords in $\{0,1\}^N$.
Since the two overlaps are strictly increasing functions of each other, maximizing the one also maximizes the other.

\subsection{Upper and Lower Bounds on Average Error Probability}
The Boolean overlap to a wrong codeword follows the binomial distribution \cite{wadsworth:60}
\begin{align}
 {\text P}_l (l) &= 2^{-N} \sum\limits_{i=0}^{\left\lfloor l  \right\rfloor} 
 \binom Ni   
 = {\text B} \left( \left\lceil N-l \right\rceil,\left\lfloor l \right\rfloor +1, \frac12\right)
\end{align}
 for $l \in[0,N]$.
Defining
\begin{equation}
\ell = \max\limits_i l_i
\label{ellmax}
\end{equation}
in analogy to \eqref{rhomax}, we get
\begin{align}
 {\text P}_{\ell|\br} (\ell)  &= \left[{\text B} \left( \left\lceil {N-\ell}  \right\rceil,\left\lfloor \ell  \right\rfloor +1, \frac12\right)\right]^{2^{NR}-1}.
\label{elldis}
\end{align}
This is the CDF of the overlap of the closest wrong codeword to the given received word $\br$. Due to the symmetry of the random code construction, it does not depends on the realization of the received word $\br$.

We can model the BSC as multiplicative noise $\bz \in \{ -1,+1\}^N$ on top of the correct codeword $\bc$.
The Boolean overlap is given by 
\begin{align}
\tau = \sum\limits_{i=1}^N \frac{1+z_i}2.
\end{align}
For crossover probability $f$, we have $\Pr(z_i=-1)=f$.
The Boolean overlap to the correct codeword is thus binomially distributed with probability 
\begin{equation}
\Pr(\tau) = 
\binom N\tau
f^{{N-\tau}}  (1-f)^\tau
\end{equation}
for  $\tau \in\setZ \cap [0,N]$.

Since the distribution is discrete, we do not get the exact error probability applying \eqref{eq1}, but the following upper and lower bounds:
 \begin{align}
 P_{\text e} &> P_{\text l} = 1-\sum\limits_{i=0}^N 
 \binom Ni f^{N-i}  (1-f)^i   {\text P}_{\ell|\br}\left(i\right)\\
 P_{\text e} &< P_{\text u} = 1-\sum\limits_{i=0}^N 
 \binom Ni f^{N-i}  (1-f)^i   {\text P}_{\ell|\br}\left(i-1\right).
  \label{pebsc}
\end{align}
These bounds emerge from \eqref{eq1} as follows: Since the random variables are discrete, there is a nonzero probability that the two Boolean overlaps equal each other.
In that case, the decoder has no other choice than guessing at random among the equiprobable options.
The lower bound and upper bound assume that in case of equality, the decisions are always correct (too optimistic) and always erroneous (too pessimistic), respectively. Clearly, the pessimistic attitude is typically much closer to reality.

\subsection{Exact Average Error Probability}

The exact average block error probability is given in \cite[Th.~2]{macmullan:98} as
\begin{align}
P_{\text e} = &1-\sum\limits_{i=0}^N 
 \binom Ni f^{N-i}  (1-f)^i   \sum\limits_{j=0}^{J-1} \frac {\binom{2^{NR}-1}j}{1+j}  \times \nonumber \\
 & \times  [{\text P}_l(i)-{\text P}_l(i-1)]^j {\text P}_{l}\left(i-1\right)^{2^{NR}-1-j}.
 \label{bscexact}
 \end{align}
 for $J=2^{NR}$. For $J<2^{NR}$, \eqref{bscexact} is an upper bound, as all factors in the two sums are strictly positive. For $J=1$, the upper bound \eqref{pebsc} is recovered. In contrast to \eqref{pebsc}, \eqref{bscexact} precisely accounts for the probability of guessing correctly.

\section{Binary Erasure Channel}
\label{bec}

Consider now a BEC with i.i.d.\ binary codewords having equal probability and erasure probability $f$.
In order to streamline the presentation with respect to the previous sections, let the alphabet of the code be $\{+1,-1\}$. Let the erasure symbol be denoted by $0$, such that  we receive $\br\in\{+1,0,-1\}^N$.
For decoding, the erased components of the received word do not matter. We can remove these components from both the received word and all the codewords in the codebook and decode with these reduced versions of the received word and the codebook.

Given the number of erasures $e$, the reduced words are uniformly distributed over the corners of a hypercube in $n=N-e$ dimensions. In analogy to \eqref{antio}, the antipodal overlap becomes
\begin{equation}
\label{antiobec}
o_i = n \cos\alpha_ i.
\end{equation}
Given the received word $\br$, the antipodal overlap of the correct codeword is not random, but fixed to $n$. 
This simplifies the following derivations, as knowledge of the full conditional CDF ${\text P}_{o|\br}(.)={\text P}_{o|e}(\cdot)$ is not required. We only need the particular value 
\begin{align}
{\text P}_{o|e}(n-1) &= 1-2^{-n}
\end{align}
which is the probability that two i.i.d.\ random codewords do not overlap in exactly $n$ given components.
Defining $\omega = \max_i o_i$ in analogy to \eqref{ellmax}, we get
\begin{align}
{\text P}_{\omega|e}(n-1) &= \left[1-2^{-n}\right]^{2^{NR}-1}.
\label{becdis}
\end{align}
Since the number of not erasures positions $n$ follows a binomial distribution, we find the average block {\it erasure} probability, see also \cite[(3.69)]{polyanskiy:10a}, as
\begin{equation}
P_{\text u} 
=1-\sum\limits_{n=0}^N 
 \binom Nn f^{N-n}  (1-f)^{n}   {\text P}_{\omega|e}\left(n-1\right).
 \label{becpu}
\end{equation}
It is an upper bound for the average block {\it error} probability by the same argument that was used for the BSC: In case of identical overlap of several codewords, there is some (small) probability to guess for the correct codeword.

The exact average block error probability is given in \cite[Th.~36]{polyanskiy:10} as
\begin{align}
P_{\text e} = &1-\sum\limits_{n=0}^N 
 \binom Nn f^{N-n}  (1-f)^{n}   \sum\limits_{j=0}^{J-1} \frac {\binom{2^{NR}-1}j}{1+j}  \times \nonumber \\
 & \times  [1-{\text P}_{o|e}(n-1)]^j {\text P}_{o|e}\left(n-1\right)^{2^{NR}-1-j}.
 \label{becexact}
 \end{align}
 for $J=2^{NR}$. For $J<2^{NR}$, \eqref{becexact} is an upper bound, as all factors in the two sums are strictly positive. For $J=1$, the upper bound \eqref{becpu} is recovered. In analogy to \eqref{bscexact}, \eqref{becexact} precisely accounts for the probability of guessing correctly.

\section{Numerical Computation}
\label{NumRes}

For 64-bit (double precision) floating point arithmetic,
the numerical evaluation of \eqref{rhodis}, \eqref{gaussfinal}, and \eqref{elldis} is not straightforward, if $NR$ exceeds values around 40. The problem is as follows:
In all cases, a CDF is raised to a huge power, before it is integrated with respect to some continuous measure.
Unless the CDF is very close to one, the huge exponent will result in a tiny result that negligibly contributes to the integral.

Floating point numbers are well-suited to deal with tiny deviations from zero, but not from one. However, the one can be converted into the other by the series expansion
\begin{align}
(1-x)^a &= {\text e}^{a \ln(1-x)}
 = {\text e}^{- \sum\limits_{i=1}^\infty\frac{ax^i}i} = \prod\limits_{i=1}^\infty{\text e}^{- \frac{ax^i}i} 
\label{prod}
\end{align}
converging for $x\in [0,1)$.
We  only need this expansion to be accurate, if $x\ll1$. For $x>10^{-8}$ and $a=2^{35}$, $(1-x)^a<10^{-149}$. This is so tiny, that its contribution to the average error probability is surely negligible.
This means that any additional term in the sum in the exponent is at least eight orders of magnitude smaller than the previous one.
This means that for $i>i_0$, all factors will be numerically equal to unity and can be ignored. In all examples discussed in this section, the threshold $i_0=1$ turns out sufficient. 

Further numerical issues arise when $2^{NR}$ exceeds the largest value of the floating point representation. For 64-bit floating point numbers, this means $NR$ is at least 1024.
The author is not aware of a general cure in that case. Depending on the channel and random code ensemble, individually tailored solutions must be found.
\subsection{Uniform Spherical Ensemble}

The incomplete beta function obeys the intrarelationship
\begin{equation}
\label{betaintra}
{\text B}(a,b,x) = 1- {\text B}(b,a,1-x).
\end{equation}
So if ${\text B}(a,b,x)$ is close to one and causes numerical trouble, ${\text B}(b,a,1-x)$ is close to zero and can be calculated very accurately in floating point arithmetic.
Utilizing \eqref{betaintra} in \eqref{prod}, we find
\begin{equation}
\label{altrhodis}
{\text P}_{\varrho|\br}(\varrho) = \prod\limits_{i=1}^\infty {\text e}^{\left(1-2^{NR}\right) \left[\frac{\text{sign}\varrho}2\,{\text B}(\frac {N-1}2,\frac 12,1-\rho^2) + 1_{\rho<0}\right]^i\frac 1i}
\end{equation}
with $1_{x<0}$ equaling 1 and 0 for $x<0$ and $x\ge0$, respectively.
If \eqref{rhodis} runs into numerical trouble, \eqref{altrhodis} will not, and vice versa, as long as $2^{NR}$ can be computed.

If $2^{NR}$ cannot be computed, we resort to the upper bound (see appendix for proof)
\begin{equation}
{\text B}\left(a,\frac12,x\right) \le  \frac1{\text B (\frac12,a)} \frac {x^a}{a\sqrt{1-x}},
\end{equation}
with equality for $x=0$, that is very tight for large $N$ and $x$ not too close to unity\footnote{
Only small values of $x$ actually have a noticeable impact for calculation of block error probability.
The incomplete beta function is strictly increasing. Thus, small arguments lead to small function values. Large function values cannot compensate for the multiplication with the hugely negative factor $1-2^{NR}$ in \eqref{altrhodis}. Thus, ${\text P}_{\varrho|\br}(\varrho)$ will be very close to zero, unless the argument of the incomplete beta function is very small. Those small values of the conditional CDF hardly add up to have a significant impact on average block error probability. 
}.
The resulting product on the right hand side of 
\begin{equation}
\label{poly}
2^{NR} {\text B}\left( \frac {N-1}2,\frac 12,1-\varrho^2\right) \le  \frac{2^R \left[ 4^R (1-\varrho^2)\right]^{\frac {N-1}2} }{\varrho \frac{N}2 \,\text B (\frac12,\frac{N+1}2)}
\end{equation}
does not run into numerical exemptions for a much wider range of code rates and blocklengths.
Note that for large $NR$, ${\text{sign}}\rho$ and $1_{\rho<0}$ in \eqref{altrhodis}, as well as higher order terms of the series expansion give vanishing contributions. Thus,
\begin{equation}
\label{dexp}
{\text P}_{\varrho|\br}(\varrho) \approx \exp  \frac{2^R \left[ 4^R (1-\varrho^2)\right]^{\frac {N-1}2} }{-\varrho {N} \,\text B (\frac12,\frac{N+1}2)}.
\end{equation}
For $NR$ around 1000, this approximation leads to relative errors below $10^{-3}$ for the full range of practically interesting block error probabilities and code rates. The larger the blocklength and rate, the more accurate is this approximation. 

The numerical evaluation of the double integral in \eqref{eq25n} is straightforward.
The integration over $s$ and $x$ can be performed by Gauss-Hermite and Gauss-Laguerre quadrature, respectively, to speed up the computation time.

Alternatively, one can use the single integral in \eqref{jumpdef}.
This was the method of choice in the late 1950s when integrals were evaluated by tables, but it is not clear, whether it is preferable for evaluation on a modern computer. The noncentral $t$-distribution cannot be expressed by polynomials combined with exponentials and/or Gaussian functions. Thus, neither Gauss-Hermite nor Gauss-Laguerre quadrature are straightforward to apply. The combination of sticking to the noncentral $t$-distribution together with the numerical issues of \eqref{rhodis} for large $NR$ may explain, why previous literature considered the exact calculation of the block error probability as numerically intractable for the most interesting ranges of blocklength.

\begin{figure}
\epsfig{file=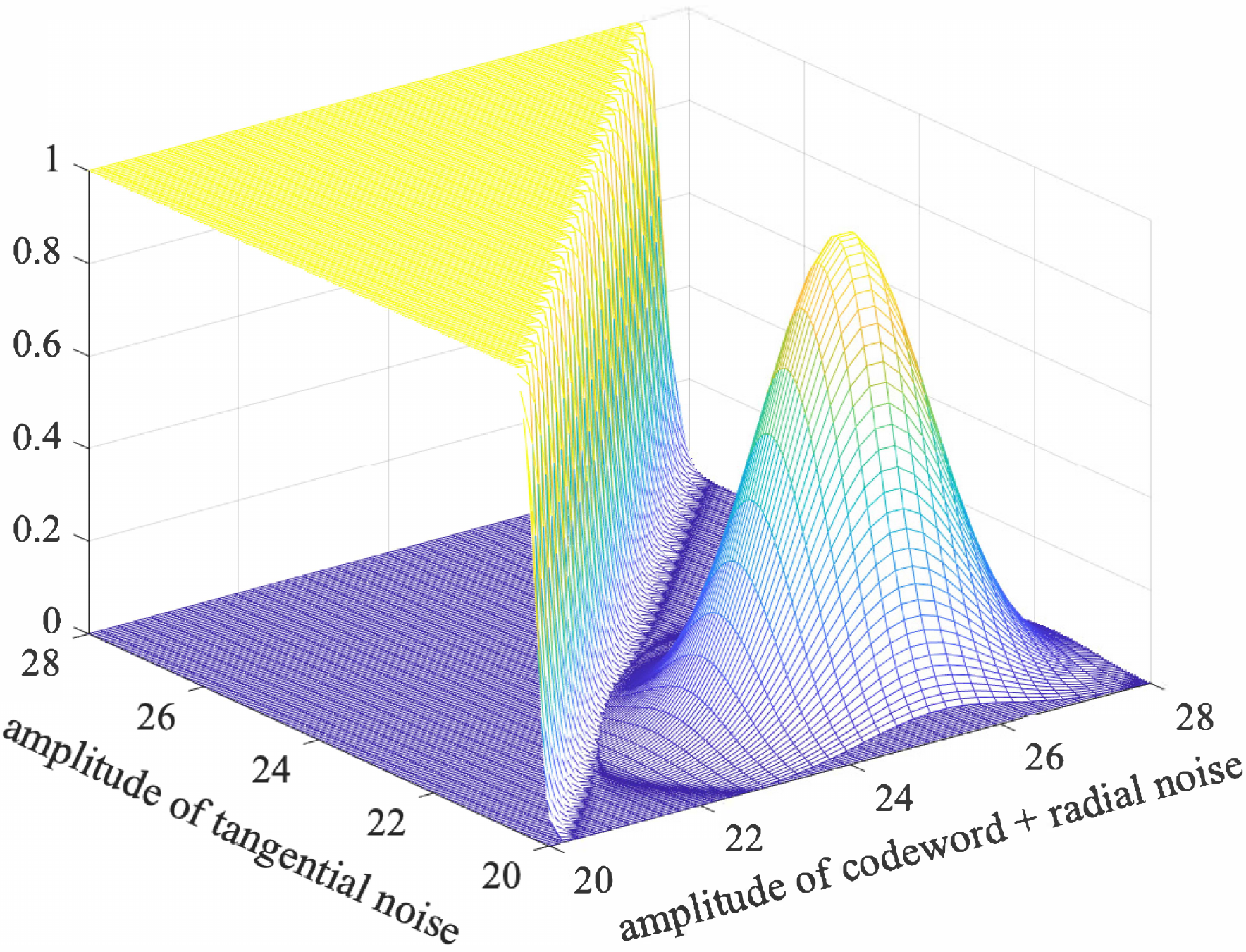,width=\columnwidth}
\caption{This figure shows 1) the joint distribution of radial and tangential components of the received word appropriately scaled to fit into the figure and 2) the probability that $\cos\beta < \varrho$, both for blocklength $N=470$, rate $R=\frac12$, signal-to-noise ratio $\log P = 1.3$ dB. The resulting block error probability is $P_{\text e} = 10^{-3}$.
\label{3D}}
\end{figure}

Figure \ref{3D} shows the joint distribution of the tangential noise $\chi$ and the sum of radial noise and codeword amplitude $s$. 
It also shows the probability $1-{\text P}_{\varrho|\br}(\cos\beta)$ with $\cos\beta$ defined in \eqref{deftan} in terms of $\chi$ and $s$.
The error wall indicates the random fluctuation of the minimum distance to other codewords.
The joint distribution shows the higher variance of the radial noise as compared to the tangential noise.
The block error probability corresponds to the integral over the product of the two.

Varying the rate in Figure \ref{3D} has the effect of changing the angle to the nearest wrong codeword.
For increasing rate the error wall turns clockwise around the origin of the coordinate system.
For low rate, the radial noise is the dominant cause of errors. For high rate, it is the tangential noise.

The conditional CDF \eqref{rhodis} scales doubly exponential with the blocklength $N$. Thus, the maximum correlation hardens much faster than the noise. This effect, clearly observed in Figure \ref{3D}, helps to explain why both the median bound and the sphere packing bound are very tight. 

\begin{figure}
\epsfig{file=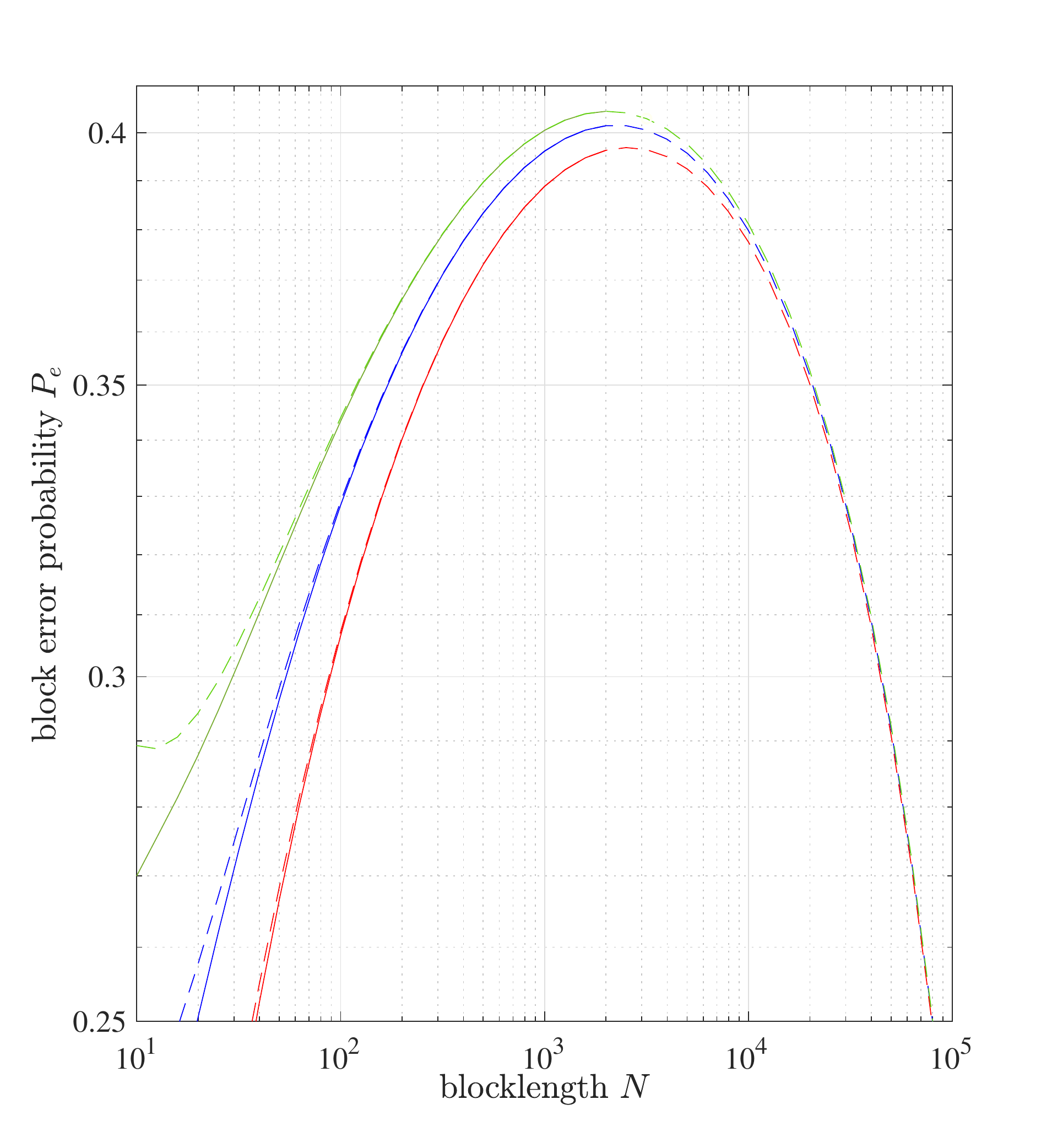,width=\columnwidth}
\caption{Block error probability vs.\ blocklength for rate $R=0.498$ and 
$P=1$.
Green, blue, and red lines refer to exact ensemble-averaged block error probability, the median bound, and the sphere packing bound. Solid lines refer to exact calculations using the incomplete beta function, dashed lines refer to approximation \eqref{dexp}.
\label{errprob}}
\end{figure}

In the sequel, we study the accuracy of approximation \eqref{dexp}. At the same time, we compare the exact calculation of the error probability to the two bounds derived from the hard error wall, i.e., the median bound and the sphere packing bound.
Figure \ref{errprob}
\label{numres}
compares the three over a wide range of blocklengths for a rate only 0.4\% below channel capacity.
The median bound is observed to lie pretty much in the middle between the exact result and the sphere packing bound.
Note that the exact calculation becomes numerically troublesome, if the product of rate and blocklength exceeds $10^3$.
For larger blocklengths, only approximation \eqref{dexp} is used. It can be observed that this approximation is sufficiently tight for practical use even for much lower blocklengths.

In order to utilize approximation \eqref{dexp} for the two bounds, one needs to solve \eqref{dexp} for $\varrho$. This cannot be done in closed form, but very efficiently by fixed point iteration:
Solve \eqref{dexp} for the $\varrho$ in the numerator (on the right hand side) assuming a fixed $\varrho$ in the denominator. Then, start the fixed point iteration for some given value of the $\varrho$ that was assumed fixed.
It is shown in the appendix that $N<2^{NR}$ is a sufficient condition for the fixed point iterations to converge.
Note that the tightness of the approximation primarily depends on the product $NR$.
For larger rates, the approximation becomes tight at even smaller blocklengths.

For rates further away from capacity, the behavior is very similar, but more difficult to depict, as the error probabilities span a very wide range in logarithmic scale. 
\begin{figure}
\epsfig{file=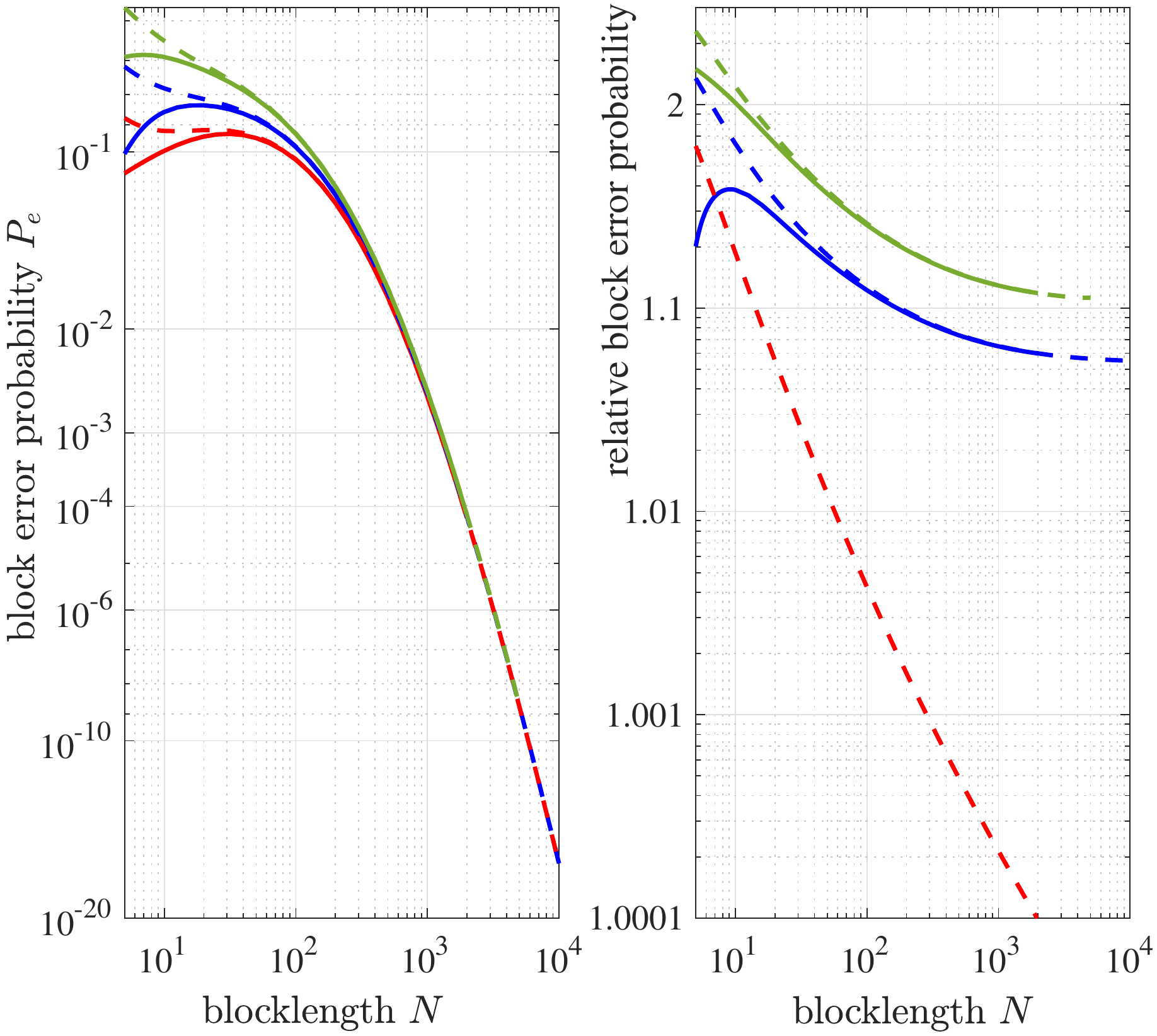,width=\columnwidth}
\caption{Absolute and relative block error probabilities vs.\ blocklength for rate $R=0.43$ and
$P=1$. Colors have the same meanings as in Fig.~\ref{errprob}. In the right figure, the block error probability is normalized by the lowest of the curves shown in the left figure.
\label{lowerror}}
\end{figure}
Absolute and relative block error probabilities are shown in Fig.~\ref{lowerror} for a rate 14\% below channel capacity. 
In order to show the large range of error probability, two subfigures are given. On the right hand side, the vertical axis is shown in nested logarithmic scale. On the left hand side the block error probability is normalized to the 1959 SPB\footnote{In order to calculate the SFB for blocklengths beyond 2000, approximation \eqref{dexp} was used.}.
For $N=10^4$, the average block error probability of the uniform spherical ensemble is only about 10 \% above the 1959 SPB. The median bound is again pretty much in the middle between the two: approximately 5\% above the 1959 SFB.

\subsection{I.i.d.\ Gaussian Ensemble}
\label{numresIID}

Depending on the number of codewords $2^{NR}$, the conditional CDF can either be calculated directly or utilizing \eqref{prod}.
The triple integral in \eqref{gaussfinal} is more time consuming than the double integral of the uniform spherical ensemble, but its evaluation can also be sped up by means of Gauss-Hermite and Gauss-Laguerre quadrature.

\begin{figure}
\epsfig{file=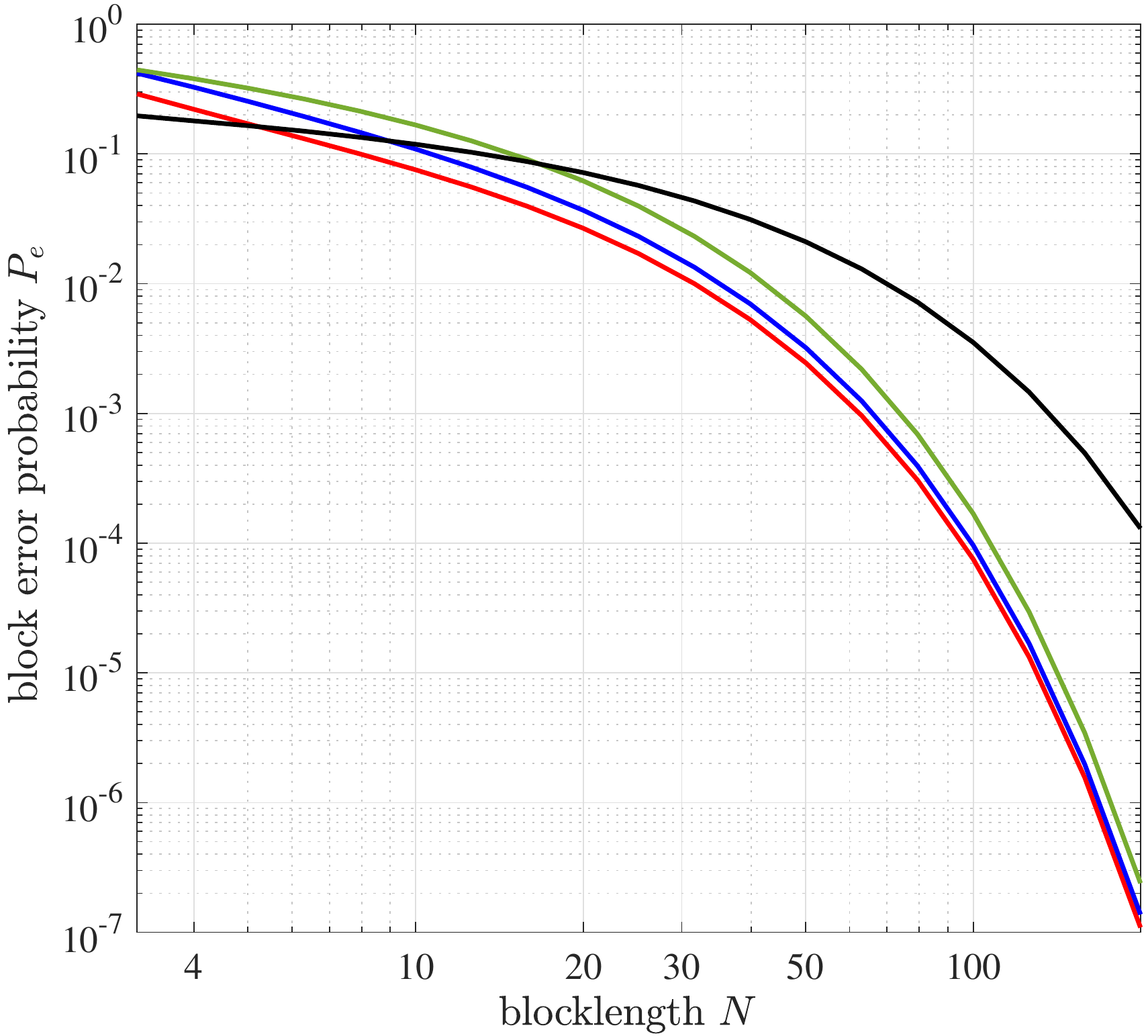, width=\columnwidth}
\caption{Block error probability vs.\ blocklength for the i.i.d.\ Gaussian ensemble (black line) for $P=100$ and $R=3$. The colored lines refer to the uniform spherical ensemble (colors as in Figure \ref{errprob}).
\label{gauss}}
\end{figure}

The i.i.d.\ Gaussian ensemble is compared against the uniform spherical ensemble in terms of average block error probability vs.\ blocklength, in Figure \ref{gauss}.
For blocklengths below 17, the i.i.d.\ Gaussian ensemble performs superior. This effect is the more pronounced the larger is the code rate.
For low dimensions, spherical codes suffer from not utilizing the radial component for data transmission. For large dimensions, this rate loss is negligible. i.i.d.\ Gaussian random codes, however, put too much information into the radial component, which is more than twice as noisy as the tangential ones.

For short blocklength, the i.i.d.\ Gaussian ensemble achieves even lower average block error probability than the sphere packing bound. This is not a contradiction. The 1959 sphere packing bound is a lower bound for best code on the hypersphere, but not for the best code overall. 

\subsection{Binary Symmetric Channel}

According to \cite{polyanskiy:10}, ''{\it the exact evaluation of \eqref{bscexact} poses considerable difficulties unless the blocklength is small.}`` In the sequel, we will explain, how to overcome these difficulties.

We start with the same strategy as for spherical and Gaussian codes via \eqref{prod} and \eqref{betaintra}. In analogy to \eqref{altrhodis}, this leads to
\begin{equation}
\label{altelldis}
{\text P}_{l} (l)^M  =\prod\limits_{i=1}^\infty {\text e}^{ -M {\text B} \left(\left\lfloor l  \right\rfloor +1,  \left\lceil N-l  \right\rceil,\frac12\right)^i \frac 1i }.
\end{equation}
%
For $NR$ exceeding 1023, the term $M\approx 2^{NR}$ in \eqref{altelldis} will cause a numerical overflow for 64-bit floating point arithmetic. This issue can be resolved as follows:
In the definition of the incomplete beta function \eqref{defbetainc}, we replace the factor $(1-\xi)^{b-1}$ by its Taylor series at $\xi=x$.
Restricting the Taylor series to first order and taking only the first factor in \eqref{altelldis}, we obtain an approximation that is sufficiently accurate for most practical cases. It reads
\begin{equation}
\label{bexp}
{\text P}_{l}(l)^M  \approx  \exp \left[{\color{red}2^{N(1-R)}{\text B}(N-l,l+1)}
\left(\textstyle
\frac{N-l-1}{l+2} - \frac{N-l}{l+1}
\right)
\right]
\vphantom{\Bigg(}
\end{equation}
assuming $l$ being an integer within $[0,N]$. 
Direct calculation of the term $2^{N(1-R)}$ may still cause a numerical overflow. This can be avoided, however, if the beta function is calculated in the logarithmic domain and combined with the exponent $N(1-R)$ prior to exponentiation.

For large blocklengths, the binomial coefficients in \eqref{bscexact} become very large, while the powers involving the crossover probabilities may become very small.
The resulting numerical trouble is overcome by calculating their product in the logarithmic domain.

Finally, \eqref{bscexact} involves a summation over $2^{NR}$ terms. However, the summation can be stopped after a few terms, without compromising the accuracy of computation. This becomes clear by recapturing the meaning of the $j^{\text{th}}$ term in the sum: Up to the factor $1/(1+j)$, it is the probability that $j$ incorrect codewords have exactly the same metric as the correct codeword. So, if we cut the series at $J$ terms, we neglect the probability that more than $J$ codewords have exactly the same metric as the correct codeword.
Clearly, $J$ need not be exponentially large for that probability to be negligible. In the examples shown below, $J\approx 10$ turned out more than sufficient.

\begin{figure}
\epsfig{file=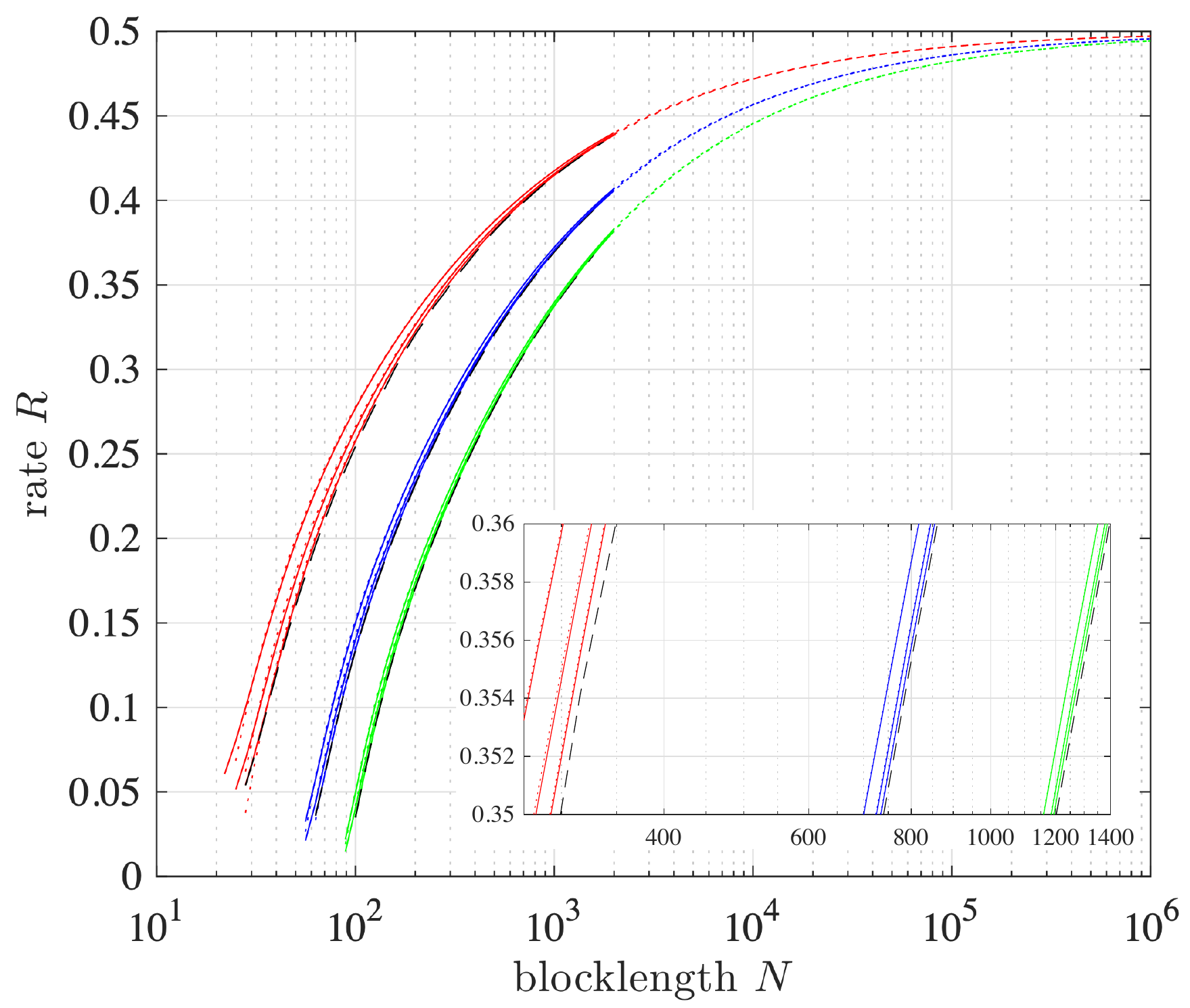,width=\columnwidth}
\caption{Rate vs.\ blocklength trade-off for the i.i.d.\ ensemble on the BSC with crossover probability $f=0.11$ for block error rates $10^{-3}$ (red), $10^{-6}$ (blue), and $10^{-9}$ (green) calculated as $P_{\text u}$, $P_{\text e}$, and $P_{\text l}$. The solid lines refer to the exact CDFs while the dotted lines show approximation \eqref{bexp}.
Dashed black lines show the respective random coding union bounds \cite[Th.~33]{polyanskiy:10}.
\label{ratevsN}}
\end{figure}

The trade-off between rate and blocklength is depicted in Figure \ref{ratevsN}.
Apparently, approximation \eqref{bexp} is very tight for the full range of blocklengths. The upper bound $P_{\text u}$ is close to the exact average block error probability.
The random coding union bound \cite{polyanskiy:10} falls only slightly behind the upper bound in \eqref{pebsc}.
Comparisons to other bounds can easily be made with the help of  \cite[Figs.~1 \& 2]{polyanskiy:10}. 

\subsection{Binary Erasure Channel}
\begin{figure}
\epsfig{file=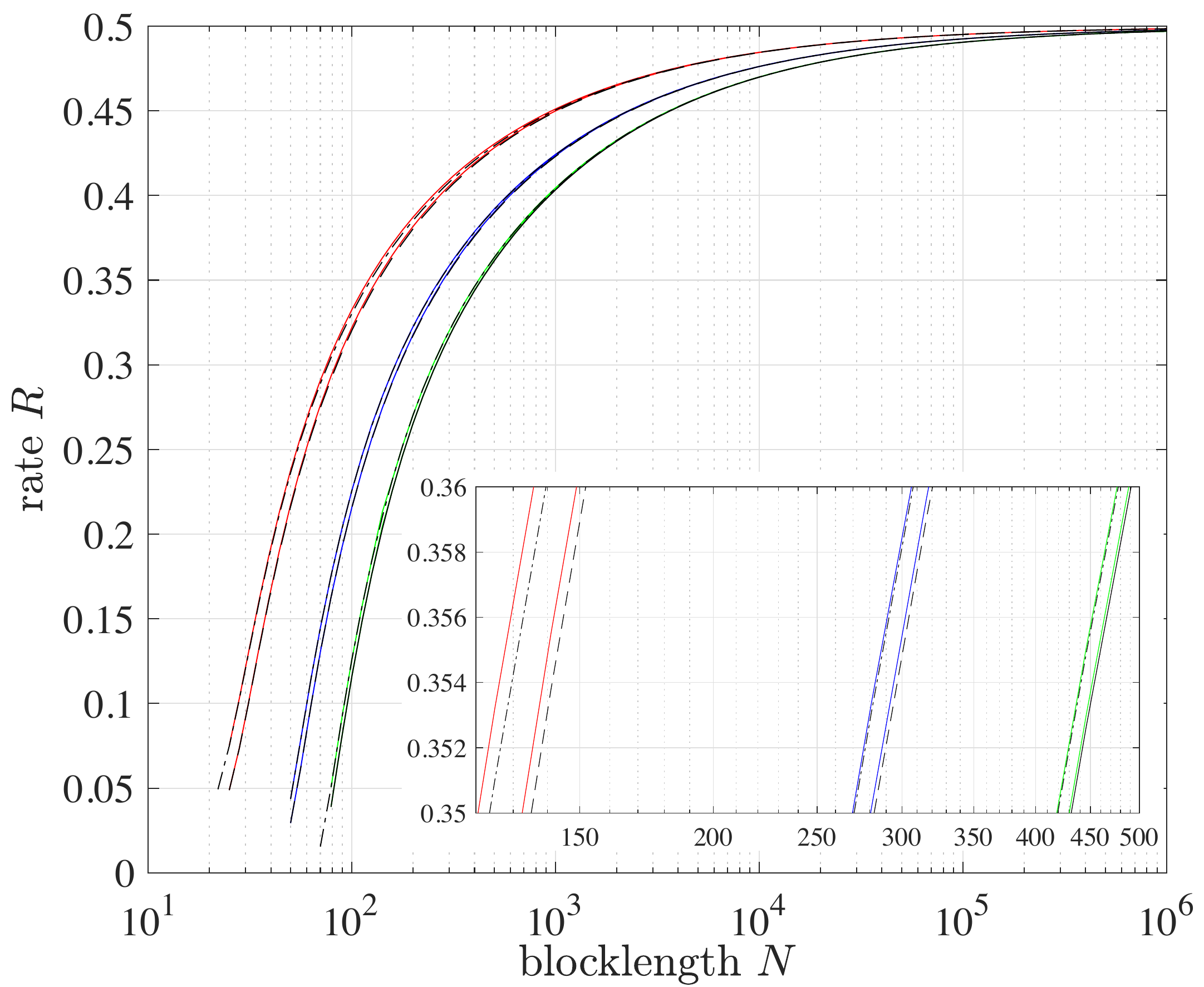,width=\columnwidth}
\caption{Rate vs.\ blocklength trade-off for the i.i.d.\ ensemble on the BEC with erasure probability $f=\frac12$ for block error rates $10^{-3}$ (red), $10^{-6}$ (blue), and $10^{-9}$ (green) claculated as $P_{\text u}$ and $P_{\text e}$. Dashed and dashed-dotted black lines show the random coding union bounds \cite[(3.61)]{polyanskiy:10a} and dependence testing bounds \cite[Th.~37]{polyanskiy:10}, respectively. 
\label{figbec}}
\end{figure}

For large values of $NR$, direct evaluation of \eqref{becdis} can run into numerical trouble.
As for the BSC, this is overcome via \eqref{prod}. In analogy to \eqref{altelldis}, this leads to
\begin{equation}
{\text P}_{o|e} (n-1)^M  =\prod\limits_{i=1}^\infty {\text e}^{ -\frac Mi 2^{in} }.
\end{equation}
The trade-off between rate and blocklength is depicted in Figure \ref{figbec}.
The random coding union bound \cite{polyanskiy:10} falls only slightly behind the upper bound in \eqref{becpu}. The dependence testing bound \cite{polyanskiy:10} is very close to the exact average error probability. Comparisons to other bounds can easily be made with the help of  \cite[Fig.~3 \& 4]{polyanskiy:10}. 

\section{Conclusions}
\label{conc}
Shannon's 1959 formula for the average block error probability of the uniform spherical ensemble can be evaluated exactly for rate-blocklength products up to 1023 with 64-bit floating point arithmetic. The same holds for i.i.d.\ random coding on both the BSC and the BEC. For larger rate-blocklength products, the average block error probability as well as Shannon's 1959 sphere packing bound can be approximated with relative errors on the order of $10^{-3}$ or less. For the uniform spherical ensemble, a tighter bound can be derived by approximating the error wall by a unit step function at its median. If the blocklength is sufficiently short and the rate sufficiently high, the uniform spherical ensemble falls behind the i.i.d.\ Gaussian ensemble.
 
\section*{Acknowledgment}
The author would like to thank S.~Asaad and A.~Bereyhi for proofreading of the manuscript, M.~Bossert, G.~Caire, A.~Guill\'en i F\`abregas, G.~Kramer, and G.~Liva for stimulating discussions, M.~Co\c{s}kun for help with some of the figures, as well as the anonynomous reviewers for pointing to many more references and suggesting an improved structure of the manuscript.
 
 \section*{Appendix}
 \subsection*{Bound on Incomplete Beta Function}
 The proof directly follows from the fact that the function $(1-\xi)^{-\frac12}$ is increasing within the unit interval.
 \begin{align}
\text B \left(a,\frac12\right) & {\text B}\left(a,\frac12,x\right)  = \int\limits_0^x \xi^{a-1} (1-\xi)^{-\frac12} {\text d}\xi\\
& \le  \int\limits_0^x \xi^{a-1} (1-x)^{-\frac12} {\text d}\xi
= \frac {x^a}{a\sqrt{1-x}}.
\label{eq40}
\end{align}
Note that by similar methods also a tight lower bound can be found, as the function $(1-\xi)^{-\frac12}$ is not only increasing, but also convex. Thus, it can be lower bounded by its first order Taylor series at $\xi=x$ which allows for integration in closed form. The resulting expression is straightforward to derive, however, it is less compact than \eqref{eq40}. In the light of the accuracy of \eqref{dexp}, we leave this as an exercise to the interested reader.

\subsection*{Convergence of Fixed Point Iteration}
For any $\varrho \in[0,1]$, 
the fixed point equation takes the form $\varrho\mapsto f(\varrho)$ with 
\begin{equation}
f(\varrho) = \sqrt{1- \left[qN\varrho \, 2^{-RN}\,{\text B}\left(\textstyle\frac12,\textstyle \frac{N+1}2\right)\right]^{\frac2{N-1}}  }
 \end{equation}
 and $q=1$ or $q=\ln 2$ for the sphere packing bound and the median bound, respectively.
 
 By Banach's fixed-point theorem \cite{subrahmanyam:18}, a necessary condition for iterations to converge is  that $\forall x,y\in[0,1]$
 \begin{equation}
 \label{banach}
 \exists\ k\in[0,1) \colon  \left|f^2(x)-f^2(y)\right| \le k \left|x^2 -y^2\right|  .
 \end{equation}
 We have
 \begin{align}
 \left|f^2(x)-f^2(y)\right| &< \left[N \,2^{-RN}\right]^{\frac2{N-1}} \left|x^{\frac2{N-1}}-y^{\frac2{N-1}}\right| \\
 &\le \left[N \, 2^{-RN}\right]^{\frac2{N-1}} \left|x^2-y^2\right|
 \end{align}
 since the beta function is smaller than 1 and $N\ge2$.
 Thus, if $N2^{-RN}<1$, \eqref{banach} is fulfilled.

\bibliography{lit}
\bibliographystyle{IEEEtran}
 
\end{document}